\documentclass[
  aps,
  pra,
  reprint,
  superscriptaddress,
  nofootinbib
]{revtex4-2} 

\usepackage[british]{babel}

\usepackage[T1]{fontenc}
\usepackage{libertinus}
\usepackage{libertinust1math}

\usepackage{mathtools}
\usepackage{amssymb}
\usepackage{siunitx}

\usepackage{graphicx}
\usepackage{tikz}
\usepackage[dvipsnames]{xcolor}

\usepackage[
  breaklinks=true,
  colorlinks=true,
  linkcolor=myblue,
  urlcolor=myblue,
  citecolor=myblue
]{hyperref}
\usepackage{orcidlink}
\usepackage{cleveref}

\makeatletter
\newcommand*{\transpose}{%
  {\mathpalette\@transpose{}}%
}
\newcommand*{\@transpose}[2]{%
  \raisebox{\depth}{$\m@th#1\intercal$}%
}
\makeatother

\newcommand{\bs}[1]{\boldsymbol{#1}}
\newcommand{\dd}{\mathop{}\!\mathrm{d}}
\newcommand{\ket}[1]{\lvert #1\rangle}
\newcommand{\bra}[1]{\langle #1\rvert}
\newcommand{\ketbra}[2]{\lvert #1\rangle\!\langle #2\rvert}

\definecolor{myblue}{RGB}{34,31,150}
\definecolor{mygreen}{RGB}{34,150,31}
\definecolor{myred}{RGB}{190,30,30}
 
\graphicspath{{figures/}}

\begin{document}

\title{Simple design of adaptive protocols for quantum metrology with limited data}

\author{Ricard Ravell Rodríguez\,\orcidlink{0000-0002-2057-5704}} 
\email{ricard.ra.ro@gmail.com}
\affiliation{ICFO-Institut de Ciències Fotòniques, The Barcelona Institute of
Science and Technology, 08860 Castelldefels (Barcelona), Spain}

\author{Jes\'{u}s Rubio\,\orcidlink{0000-0002-8193-8273}}
\email{jesus@rubiojimenez.com}
\affiliation{Department of Physics and Astronomy, University of Exeter, Stocker Road, Exeter EX4 4QL, United Kingdom}

\date{\today}

\begin{abstract}
Recent sensing experiments have used a Bayesian notion of precision gain to adapt control parameters for fixed measurements.
Here, we advocate a simple design principle based on Personick's minimum mean square loss formalism to extend this procedure in order to include measurement optimisation, determining the optimal measurement for each probe and providing a prior-dependent precision gain to optimise the probe itself.
This offers a Bayesian counterpart to protocol design based on the Fisher information, with a similar algebraic structure but incorporating the full prior.
We apply this principle to adaptive phase estimation with single-mode Gaussian probes, showing that joint optimisation outperforms both adaptive homodyne detection and a fixed-probe strategy optimised for the initial prior, with a maximum reduction of $32\%$ in mean posterior variance relative to homodyne detection.
Our results illustrate the potential of simple Bayesian design for quantum metrology with limited data when measurement bases can be controlled.
\end{abstract}

\maketitle

\section{Introduction}

\begin{table}[t]
\caption{Parallel between local (SLD) and Bayesian (SPM, single shot) probe and measurement design. The local framework follows Refs.~\cite{helstrom1976quantuum,braunstein1994statistical,DemkowiczDobrzanski2015}, while the Bayesian framework is discussed in Secs.~II and III.}
\label{tab:sld-spm}
\begin{ruledtabular}
\begin{tabular}{lcc}
 & Local (SLD) & Bayesian (SPM) \\
\hline
Prior information & localisation at $\theta_0$ & arbitrary $p(\theta)$ \\
Operator equation & $\rho_{\theta_0} L+L\rho_{\theta_0}=2\partial_\theta\rho_\theta|_{\theta_0}$ & $\rho_0 S+S\rho_0=2\bar\rho$ \\
Optimal measurement & eigenbasis of $L$ & eigenbasis of $S$ \\
Figure of merit & $\mathcal I=\mathrm{Tr}(\rho_{\theta_0} L^2)$ & $\tilde{\mathcal G}=\mathrm{Tr}(\rho_0 S^2)$ \\
Precision bound & $\mathrm{Var}\ge 1/(\mu\mathcal I)$ & $\mathcal L_{\mathrm{MSL}}\ge\lambda-\mathrm{Tr}(\rho_0 S^2)$ \\
Probe design & $\max_{\bs y}\mathcal I$ & $\max_{\bs y}\tilde{\mathcal G}$ \\
\end{tabular}
\end{ruledtabular}
\end{table}

Quantum metrology has traditionally focused on maximising the Fisher information (FI)~\cite{DemkowiczDobrzanski2015}.
The FI quantifies the local distinguishability of neighbouring states and determines Cramér--Rao (CR) precision bounds~\cite{helstrom1976quantuum, braunstein1994statistical}.
Yet, local distinguishability does not always determine how well a parameter can be estimated from a finite number of observations.
In phase estimation, for instance, states with a large FI can produce several indistinguishable candidate phases, as occurs with N00N states, making their apparent sensitivity difficult to exploit without sufficient prior information~\cite{Durkin2007,
berry2009accurate,hayashi2011phase}.
This problem can become particularly acute when the available data are not only finite but also \emph{limited}~\cite{lumino2017experimental,valeri2020experimental,Rubio_2019,rubio2020thesis}.

Related difficulties emerged with protocols apparently offering sub-Heisenberg precision~\cite{rivas2012sub}. 
These issues motivated non-asymptotic analyses of precision limits accounting for finite prior information~\cite{hall2012universality,hall2012nonlinear,berry2012heisenberg,tsang2012ziv,luis2017breaking}, as well as of the actual precision of FI-optimised protocols and the number of repetitions required to approach their asymptotic regime~\cite{rubio2018non}.
The validity of asymptotic approximations had in fact been examined as early as the 1990s~\cite{braunstein1992how}.
A complementary explanation was later found in the inadequacy of average photon number as a resource metric, which can obscure the photons responsible for the sensitivity and suggest spurious sub-Heisenberg scaling~\cite{branford2021average, izquierdo2026precision}.
Renewed interest in the limitations of the FI has also led to revisiting some of these matters~\cite{hradil2026realistic}.

Several approaches to metrology with finite data have been developed.
One line of research has sought precision bounds beyond the CR bound. 
These include the Bay\-es\-ian Ziv-Za\-kai~\cite{tsang2012ziv}, Weiss-Wein\-stein~\cite{lu2016weiss}, and van Trees~\cite{martinezvargas2020quantum} bounds, as well as frequentist approaches based on the Bhat\-ta\-char\-yya and Ba\-ran\-kin bounds~\cite{gessner2023hierarchies,gebhart2024fundamental,navarro2025existence,reichert2026quantumbarankinboundslocal} and other generalisations of the CR bound~\cite{garciamartinez2026squeezing}.
Finite-sample precision has also been characterised via the probability of estimating a parameter within a given error, with bounds derived from hypothesis testing in Ref.~\cite{meyer2025finite}; this work provides a useful overview of a range of approaches to finite-data quantum metrology.

A different direction has focused on optimising estimation protocols directly for the available data.
Bay\-es\-ian inference is naturally suited to this task, as it incorporates prior information from the outset, provides optimal estimators directly, and gives probabilities a straightforward operational meaning even at the single-shot level~\cite{jaynes2003probability}---which is needed when only limited data are available.

In this context, Personick determined the single-shot minimum mean square loss (MSL) and the operator whose eigenbasis defines the optimal measurement~\cite{Personick,helstrom1976quantuum}, sometimes referred to as the symmetric posterior mean (SPM) in modern literature~\cite{albarelli2025measurement}.
Ref.~\cite{Macieszczak_2014} applied this result to the joint optimisation of probe and measurement in frequency estimation.
It also introduced two key insights: the use of projective measurements does \emph{not} lead to a loss of optimality---a step missing in Personick's original derivation---and, for Gaussian priors, the minimum MSL can be expressed in terms of the quantum FI of a prior-averaged state.
A related connection between minimum MSL and FI can likewise be established by linearising the state around the prior mean \cite{branford2021average}.

This framework was then applied to a finite number of individual observations via repetitions of interferometric runs, revealing differences in the relative performance of quantum probes not captured by the quantum FI~\cite{Rubio_2019, nichols2019designing}.
It was also generalised to accommodate loss functions informed by parameter symmetries and information geometry~\cite{boeyens2025quantum}, with methods for closed-form optimisation under investigation \cite{gandar2026closedformbayesianquantumestimation}. 
Together with recent numerical advances based on semidefinite programming~\cite{bavaresco2024designing}, which allow the optimisation of probes, measurements, and estimators for arbitrary loss functions, these approaches cover a wide range of finite-data metrology problems that can be constructively solved.

Of particular relevance are adaptive strategies~\cite{BerryWiseman2000,BerryWisemanBreslin2001,rodriguezgarcia2024adaptive, ricard-simon, andre2026strategy}, in which previous measurement outcomes guide subsequent design choices.
These have been experimentally implemented, e.g., in optical phase estimation~\cite{Paesani_2017,lumino2017experimental,valeri2020experimental} and single-ion microscopy using the Kullback--Leibler divergence as the information metric~\cite{jacob2016microscopy,jacob2016maximizing}.
More recently, cold atom experiments have aimed to set the value of control parameters by maximising a Bayesian notion of precision gain defined in terms of a given measurement scheme~\cite{glatthard2022optimal,overton2026adaptive,fess2026bayesian}.
These developments motivate the use of greedy adaptive strategies, in which each experimental run is treated as the unit of optimisation.
Although such strategies need not be globally optimal~\cite{andre2026strategy}, they provide a practical approach to experimental design when only limited data are available.

Given this state of affairs, a compromise between the flexibility of direct optimisation and the computational complexity of general numerical methods could be desirable for experiments in which measurement bases can be adjusted.
Personick's framework provides a simple design principle, yielding an analytically defined optimal measurement and a Bayesian precision gain that can be maximised over probe parameters.
Methodologically, this parallels maximising the quantum FI and measuring in the eigenbasis of the symmetric logarithmic derivative (SLD) in local estimation~\cite{helstrom1976quantuum,braunstein1994statistical} (Table~\ref{tab:sld-spm}), while incorporating arbitrary prior information.

Here, we construct greedy adaptive protocols by exploiting this parallel, extending the experimental optimisation of control parameters for fixed measurements in Refs.~\cite{glatthard2022optimal,overton2026adaptive,fess2026bayesian}.
Specifically, we develop a simple adaptive protocol for phase estimation with single-mode Gaussian probes~\cite{Morelli_2021}, in which both the probe and its optimal quantum measurement are reoptimised after each observation.
At each round, the posterior, a distribution that incorporates the information gained from previous measurement outcomes, serves as the prior for the next run.
The performance of the complete experiment is assessed through its MSL, evaluated using the optimal Bayes estimator conditioned on all measurement outcomes.
We show that joint optimisation yields the lowest estimation uncertainty among the strategies considered, achieving a maximum relative improvement of approximately $32\%$ over adaptive homodyne detection at $17$ measurements, with a corresponding improvement of $50\%$ over the fixed-probe strategy.
Our study thus provides a concrete illustration of how the proposed Bayesian design principle can be implemented in a standard quantum metrology setting.

\section{Partially informed phase estimation}

In this section, we outline a simple phase estimation protocol serving as a testbed for the adaptive strategies discussed in Sec.~\ref{sec:adaptive-estimation}.

\subsection{Protocol for a single shot}
\label{sec:protocol}

Let \(\Theta\) be an unknown parameter and \(\theta \in [\theta_\mathrm{min}, \theta_\mathrm{max}]\) a candidate value for it.
We shall refer to \(\theta\) as a \emph{hypothesis} for \(\Theta\)~\cite{jaynes2003probability}.
The hypothesis is assumed to be encoded into an initial pure probe state \(\ket{\psi_{\bs{y}}^{\mathrm{in}}}\) through the unitary transformation
\begin{equation}
    \ket{\psi_{\bs{y}}(\theta)} 
    = \mathrm{e}^{-i \theta \hat{a}^\dagger \hat{a}} \ket{\psi_{\bs{y}}^{\mathrm{in}}}.
    \label{eq:transformed-state}
\end{equation}
Here, \(\hat{a}\) and \(\hat{a}^\dagger\) denote the annihilation and creation operators of a bosonic system, and \(\bs{y}\) is a vector of parameters defining the family of initial states. 
We call \(\bs{y}\) control parameters---these will later enable probe optimisation.

The state \(\ket{\psi_{\bs{y}}^{\mathrm{in}}}\) is further assumed to be Gaussian.
Accordingly, the parametrised state in Eq.~\eqref{eq:transformed-state} is fully characterised by the vector of first moments~\cite{loudon2000quantum,braunstein2005quantum,weedbrook2012gaussian,serafini2017quantum,brask2022gaussianstatesoperations}
\begin{align}
    \overline{\mathbf{r}}
    = \sqrt{2}|\alpha|
    \begin{pmatrix}
        \cos(\tau-\theta) \\
        \sin(\tau-\theta)
    \end{pmatrix}
    \label{eq:first moment single mode states_2}
\end{align}
and the covariance matrix
\begin{align}
    V 
    = \begin{pmatrix}
        V_{qq} & V_{qp} \\
        V_{qp} & V_{pp}
    \end{pmatrix},
    \label{eq:cov matrix single mode states_2}
\end{align}
where the components are given by
\begin{equation}
\begin{aligned}
    2 V_{qq\,(pp)}
    &= \left[
        \cosh(2r)
        \mp
        \cos(2\theta-\varphi)\sinh(2r)
    \right],
    \\
    2 V_{qp}
    &= \sin(2\theta-\varphi)\sinh(2r),
    \label{eq:covariance-componentes}
\end{aligned}
\end{equation}
with the upper (lower) sign corresponding to \(V_{qq}\) (\(V_{pp}\)).
A brief review of Gaussian states is provided in App.~\ref{app:gaussian-probes}.

Within this state family, the displacement parameter is \(\alpha = |\alpha| e^{i\tau}\), with \(\tau \in [0,2\pi)\), and the squeezing parameter is \(\xi = r e^{i\varphi}\), with \(\varphi \in [0,2\pi)\).
We also define \(\bs{y} \coloneq (|\alpha|,\tau,\varphi)\), while we fix \(r\) through
\begin{equation}
    r = \mathrm{arsinh}\big(\sqrt{E-|\alpha|^2}\big),
    \label{eq:energy-constraint}
\end{equation}
where $E$ denotes the probe energy, assumed to be known and fixed, following Ref.~\cite{ricard-simon}, and $|\alpha| \in [0, \sqrt{E}]$.
In our numerical calculations, we choose \(E=2\), which suffices to produce nontrivial results while keeping the Fock-space cutoff low \cite{Rubio_2019,ricard-simon}.

To extract information about \(\Theta\), a measurement is performed on \(\ket{\psi_{\bs{y}}(\theta)} \). 
The operation is described by a positive operator-valued measure (POVM), denoted by \(\{M(x)\}\), with outcomes \(\{x\}\). 
The probability of obtaining a particular outcome is given by \(p(x|\theta,  \boldsymbol{y}) = 
\bra{\psi_{\bs{y}}(\theta)} M(x)\ket{\psi_{\bs{y}}(\theta)}\)---the Born rule.
Then, if we start with a prior distribution \(p(\theta)\) encoding the \emph{a priori} information (or lack thereof) about \(\Theta\), and we \emph{update} it with the information provided by a single measurement run, the overall information about \(\Theta\) is given by the posterior distribution
\begin{equation}
    \label{eq:posterior}
    p(\theta | x, \boldsymbol{y})
    \propto
    p(\theta)\,
    \bra{\psi_{\bs{y}}(\theta)} M(x)\ket{\psi_{\bs{y}}(\theta)},
\end{equation}
according to Bayes' theorem.

A point estimator \(\tilde{\theta}(x)\), mapping the outcome \(x\) to an estimate for \(\Theta\), may be defined as a functional of the posterior distribution in Eq.~\eqref{eq:posterior} \cite{jaynes2003probability, DemkowiczDobrzanski2015}.
This estimator represents our best-informed guess of the true value and is, therefore, the ultimate output of the protocol, notwithstanding the need to calculate, in addition, outcome-dependent uncertainty quantifiers such as error bars or credible intervals \cite{jaynes1976confidence, boeyens2021uninformed, boeyens2025quantum} that we do not consider in this work.

\subsection{Optimality criterion}

An immediate question is how to choose the POVM \(M(x)\) and estimator \(\tilde{\theta}(x)\), as well as how to tune the control parameters \(\bs{y}\).
The answer is to formulate an optimisation problem over a suitable figure of merit.

Since protocol design precedes data acquisition, such a figure of merit cannot depend on either the measurement outcome or a particular hypothesis for the unknown value of $\Theta$ \cite{rubio2020thesis}. 
We thus consider an uncertainty quantifier averaged over both, namely the mean loss
\begin{equation}
    \mathcal{L}[\bs{y}, M(x), \tilde{\theta}(x)]
    \coloneqq \int \dd \theta \dd x\,
    p(\theta)p(x|\theta, \bs{y})
    \ell[\tilde{\theta}(x),\theta],
    \label{eq:mean-loss}
\end{equation}
where the loss function \(\ell(\tilde{\theta},\theta)\) quantifies the cost incurred by the estimator \(\tilde{\theta}\) if \(\theta\) were the true value.

The loss function must respect the symmetries of the problem~\cite{Rubio-symm}.
Since Eq.~\eqref{eq:transformed-state} is invariant under $\theta\to\theta+2\pi$, a natural choice is the sinusoidal loss
$\ell(\tilde{\theta},\theta)=4\sin^2[(\tilde{\theta}-\theta)/2]$ (see App.~\ref{app:loss-function}), whose minimum mean loss is known in closed form~\cite{demkowicz2011optimal}.
However, for the flat prior $p(\theta)=2/\pi$ on $[0,\pi/2]$ we use in this work, the sinusoidal loss can be approximated by the squared loss
$\ell(\tilde{\theta},\theta)=(\tilde{\theta}-\theta)^2$~\cite{Rubio_2019,Morelli_2021}.
Eq.~\eqref{eq:mean-loss} then becomes the familiar mean square loss (MSL).

Next, to ease the optimisation of the MSL, discussed in the next section, we rewrite it as 
\begin{equation}
    \mathcal{L}_{\mathrm{MSL}} = \lambda + \mathrm{Tr}(\rho_0 M_2 - 2 \bar{\rho} M_1),
\end{equation}
where $\lambda = \int \dd \theta p(\theta) \theta^2$. 
Here, we define the measurement $k$-th moment
\begin{equation}
    M_k \coloneqq \int \dd x\,M(x)\tilde{\theta}(x)^k
\end{equation}
and, in our specific scenario based on pure states, the state zeroth and first moments
\begin{subequations}\label{eq:rho-both}
    \begin{align}
       \rho_0
       &\coloneqq \int \dd \theta\, p(\theta)\,
       \ketbra{\psi_{\bs{y}}(\theta)}{\psi_{\bs{y}}(\theta)},
       \label{eq:rho0}
       \\
       \bar{\rho}
       &\coloneqq \int \dd \theta\, p(\theta)\,
       \ketbra{\psi_{\bs{y}}(\theta)}{\psi_{\bs{y}}(\theta)}\,\theta,
       \label{eq:rhobar}
    \end{align}
\end{subequations}
respectively.
Accordingly, we define the optimal estimation strategy as the triple \((M, \tilde{\theta}, \bs{y})\) minimising \(\mathcal{L}_{\mathrm{MSL}}\), i.e., a \emph{joint} probe and measurement optimisation within the family of Gaussian states defined in Sec.~\ref{sec:protocol}.
Note that, unlike in local estimation, pure states generally become mixed in a Bayesian setting, as per Eq.~\eqref{eq:rho0}.

The approximation of the periodic loss by the squared loss introduces an element of localisation into an otherwise global estimation problem.
Nevertheless, the optimisation remains Bayesian as the loss is still averaged over all possible hypotheses within the relevant parameter region.
We refer to this approach as \emph{partially informed} phase estimation~\cite{rubio2020thesis}.
    
\subsection{Optimal single-shot strategy}
\label{eq:optimal-1shot-strategy}

\citet{Personick} showed, with a variational calculation refined by later authors~\cite{Macieszczak_2014, rubio2023quantum}, that the minimum MSL is given as
\begin{equation}
    \mathcal{L}_{\mathrm{min}} = \lambda - \mathrm{Tr}(\rho_0 S^2)
    \label{eq:min-msl}
\end{equation}
for a single shot, where $S$ is the SPM operator \cite{albarelli2025measurement} found by solving the Sylvester equation
\begin{equation}
    S \rho_0 + \rho_0 S = 2 \bar{\rho}.
    \label{eq:Lyapunov}
\end{equation}
This result can alternatively be derived through a Bayesian analogue of the Braunstein--Caves inequality~\cite{braunstein1994statistical}, following the approach of Ref.~\cite{rubio2020bayesian}.

The optimal strategy that minimises the MSL is determined by the SPM operator, as follows.
Given its eigendecomposition
\begin{equation}
S = \int \dd s \mathcal{P}(s) s,
\end{equation}
where $\mathcal{P}(s)\mathcal{P}(s')=\delta(s-s')\mathcal{P}(s')$, the optimal POVM is obtained by projecting onto its eigenbasis, namely
\begin{equation}
    M_{\mathrm{opt}}(s)=\mathcal{P}(s).
    \label{eq:opt-povm}
\end{equation}
Furthermore, the optimal estimator is given by the associated eigenvalue, $\tilde{\theta}_{\mathrm{opt}}(s)=s$, which can be shown to be equivalent to the posterior mean estimator \cite{rubio2023quantum},
\begin{equation}
    \tilde{\theta}_{\mathrm{opt}}(s)
    = \int \dd\theta p(\theta|s,\bs{y})\theta,
    \label{eq:opt-est}
\end{equation}
which is the classical Bayes estimator under the squared loss criterion \cite{jaynes2003probability}.
This equivalence is the origin of the name SPM operator.

Eq.~\eqref{eq:min-msl} can also be written as $\mathcal{L}_{\mathrm{min}} = \sigma_0^2-\mathcal{G}$, where $\sigma_0^2=\lambda-m_0^2$ is the prior variance,
$m_0=\int\dd\theta\,p(\theta)\theta$ is the prior mean, and
\begin{equation}
 \mathcal{G}\coloneqq\mathrm{Tr}(\rho_0 S^2)-m_0^2   
\end{equation}
is the precision gain, quantifying the reduction in uncertainty provided by the optimal measurement.

Since $m_0$ is independent of the control parameters, minimising $\mathcal{L}_{\mathrm{min}}$ over the probe parameters is equivalent to maximising the \emph{pseudo}-gain~\cite{albarelli2025measurement} as 
\begin{equation}
    \tilde{\mathcal{G}}
    \coloneqq
    \mathrm{Tr}(\rho_0 S^2),
    \quad
    \bs{y}_{\mathrm{opt}}
    =
    \underset{\bs{y}}{\operatorname{argmax}}
    \,\tilde{\mathcal{G}}.
    \label{eq:opt-control}
\end{equation}
This parallels the maximisation of the quantum FI in local estimation, while incorporating the full prior.

For a fixed POVM, the analogous \emph{classical} pseudo-gain is
\begin{equation}
    \tilde{\mathcal{G}}_{\mathrm{cl}}
    \coloneqq
    \int \dd x\,p(x|\bs{y})
    \tilde{\theta}_{\mathrm{opt}}(x)^2,
    \label{eq:classical-pseudogain}
\end{equation}
where $p(x|\bs{y})=\int\dd\theta\,p(\theta)p(x|\theta,\bs{y})$
and $\tilde{\theta}_{\mathrm{opt}}(x)$ is the posterior mean in Eq.~\eqref{eq:opt-est}.
This is the quantity experimentally optimised in Refs.~\cite{glatthard2022optimal,overton2026adaptive,fess2026bayesian}, with appropriate modifications to account for the symmetries of the estimation problem at hand~\cite{Rubio-symm}.
The SPM construction instead optimises the measurement itself, in analogy with the distinction between classical and quantum FI.
This distinction was previously explored for precision gains in global thermometry and scale estimation~\cite{rubio2021global,rubio2023quantum}.

Eqs.~\eqref{eq:opt-povm}, \eqref{eq:opt-est} and \eqref{eq:opt-control} define the optimal quantum strategy for a single shot within the Bayesian framework.
In the next section, we incorporate this strategy into adaptive multishot protocols.

\section{Adaptive multishot estimation}
\label{sec:adaptive-estimation}

In this section we describe the adaptive protocols considered in this work, before presenting our results in Sec.~\ref{sec:results}.

\subsection{Optimal estimator and loss}

Suppose that \(\mu\) measurements are performed, yielding the outcome vector
\(\bs{x}=(x_1,\ldots,x_\mu)\), while allowing both the control parameters and the POVM to vary between shots.
The posterior distribution in Eq.~\eqref{eq:posterior} generalises to
\begin{equation}
    \label{eq:posterior-multishot}
    p(\theta | \bs{x}, \{\bs{y}\})
    \propto p(\theta)\prod_{i=1}^{\mu}
    \bra{\psi_{\bs{y}_i}(\theta)} M_i(x_i)\ket{\psi_{\bs{y}_i}(\theta)},
\end{equation}
where \(\{\bs{y}\}\) denotes the collection of control vectors, with $i$-th element
\(\bs{y}_i=(|\alpha|_i,\tau_i,\varphi_i)\).

For the multishot MSL, the classical Bayes estimator is the posterior mean \cite{jaynes2003probability},
\begin{equation}
    \tilde{\theta}_{\mathrm{opt}}(\bs{x})
    = \int \dd\theta
    p(\theta | \bs{x}, \{\bs{y}\})\theta,
    \label{eq:opt-est-multishot}
\end{equation}
which is the multishot generalisation of Eq.~\eqref{eq:opt-est}.
This estimator admits any sequence of POVMs and control parameter values. 
The associated MSL is the mean posterior variance (MPV)
\begin{equation}
    \mathcal{L}_{\mathrm{MPV}} = \lambda - \int \dd \bs{x} p(\bs{x}|\{\bs{y}\}) \tilde{\theta}_{\mathrm{opt}}(\bs{x})^2,
    \label{eq:mpv}
\end{equation}
where $p(\bs{x}|\{\bs{y}\}) = \int \dd \theta p(\theta) p(\bs{x} | \theta, \{\bs{y}\})$. 

We shall evaluate it numerically, with the parameter integrals computed deterministically and the outcome-related averages estimated from $1000$ Monte Carlo runs.
Our implementation can be found in Ref.~\cite{code}.

\subsection{Finite-sample strategies}
We now specify the three adaptive strategies considered in this work.
They differ in whether the measurement, the probe, or both are reoptimised after each observation.
In each case, the probe optimisation uses the appropriate classical or quantum pseudo-gain defined in Sec.~\ref{eq:optimal-1shot-strategy}.
For all of them, let
\begin{equation}
        \bs{x}_{j-1}
        \coloneq
        (x_1,\ldots,x_{j-1}),
        \quad
        \{\bs{y}\}_{j-1}
        \coloneq
        (\bs{y}_1,\ldots,\bs{y}_{j-1})
\end{equation}
and write
\begin{equation}
    p_{j-1}(\theta)
    \coloneq
    p(\theta|\bs{x}_{j-1},\{\bs{y}\}_{j-1}),
\end{equation}
with $p_0(\theta)\equiv p(\theta)$.
At measurement $j$, this distribution is used as the prior for the next optimisation.
Once the full sequence of outcomes, measurements, and control parameters has been generated, the posterior in Eq.~\eqref{eq:posterior-multishot} is used to evaluate the MPV in Eq.~\eqref{eq:mpv}.

The first strategy keeps the probe fixed and adapts the measurement.
We choose the probe by maximising the quantum pseudo-gain in Eq.~\eqref{eq:opt-control} for the initial uniform prior, under the energy constraint in Eq.~\eqref{eq:energy-constraint}, and keep it unchanged throughout the protocol.
At round $j$, we use the current prior $p_{j-1}(\theta)$ to obtain $\rho_{0,j}$ and $\bar{\rho}_j$ according to Eqs.~\eqref{eq:rho-both}, and hence the SPM operator $S_j$ from Eq.~\eqref{eq:Lyapunov}.
We then measure in its eigenbasis according to Eq.~\eqref{eq:opt-povm}, reoptimising the measurement after each observation.
This extends the shot-by-shot methodology of Ref.~\cite{Rubio_2019}, in which the optimal single-shot strategy is repeated to construct a finite-sample protocol (an approach also implemented experimentally with cold atoms~\cite{glatthard2022optimal,overton2026adaptive}).

The second strategy reverses these roles, fixing the measurement to homodyne detection while adapting the parameters controlling the shape of the Gaussian probe.
At every shot, we measure the position quadrature $q$, so that the generic POVM $M(x)$ becomes $M(q)=\ketbra{q}{q}$.
This choice is motivated by the fact that homodyne detection attains the ultimate local precision available for the Gaussian protocol described in Sec.~\ref{sec:protocol} when the squeezing angle is set at the appropriate working point~\cite{Monras_2006,Olivares_2009,Gaiba_2009}.
This precision is quantified by the quantum FI, which, for the unitary transformation of a pure state in Eq.~\eqref{eq:transformed-state}, can be calculated simply as
\begin{equation}
    \mathcal I
    =
    4[
        \langle(\hat a^\dagger\hat a)^2\rangle
        -
        \langle\hat a^\dagger\hat a\rangle^2
    ],
    \label{eq:qfi-pure}
\end{equation}
with maximum
\begin{equation}
    \mathcal I_{\max}
    =
    2\sinh^2(2r)
    =
    8E(E+1)
    \label{eq:max-qfi}
\end{equation}
attained by a squeezed-vacuum probe, $|\alpha|=0$, for which $E=\sinh^2r$.
With the measurement fixed, the probe parameters are adapted at each round by maximising the classical pseudo-gain in Eq.~\eqref{eq:classical-pseudogain}, evaluated using the current prior $p_{j-1}(\theta)$.
This extends Ref.~\cite{ricard-simon} by optimising directly over the full posterior distribution at each round, rather than relying on a Gaussian approximation and the associated high- and low-uncertainty strategies introduced there.

The third strategy adapts both the probe parameters and the measurement.
At round $j$, we use the current prior $p_{j-1}(\theta)$ to construct the SPM operator $S_j$ from Eq.~\eqref{eq:Lyapunov} for each candidate probe.
The control parameters $\bs{y}_j$ are then chosen by maximising the quantum pseudo-gain according to Eq.~\eqref{eq:opt-control}, and we measure in the eigenbasis of the corresponding SPM operator.
After each outcome, the posterior is updated and the procedure is repeated.
This combines the adaptive probe optimisation of the second strategy with the adaptive Bayesian measurement of the first.

For the two strategies involving SPM operators, the POVM is often non-Gaussian, so we introduce a finite Fock-space cutoff of $100$ in the numerical calculations.
This cutoff is reasonable for an average energy of $E=2$~\cite{rubio2018non}.
No such truncation is required for the homodyne strategy, whose statistics are fully determined by the first and second moments of the Gaussian probe.

\subsection{Relative improvement and asymptotic benchmark}

We compare finite-sample strategies through the MPV in Eq.~\eqref{eq:mpv}.
To quantify the improvement of one strategy over another for a number of measurements $\mu$, we define
\begin{equation}
    R_{A|B}(\mu)
    \coloneqq
    100\left[
    1-
    \frac{\mathcal{L}_{\mathrm{MPV}}^A(\mu)}
         {\mathcal{L}_{\mathrm{MPV}}^B(\mu)}
    \right]\%,
    \label{eq:relative-improvement}
\end{equation}
where $A$ and $B$ denote the strategy of interest and the reference strategy, respectively.
A positive value indicates a reduction in estimation uncertainty.
We shall use this quantity to identify the number of measurements at which joint optimisation achieves its maximum relative improvement over adaptive homodyne detection within the range considered.

The asymptotic error is also considered for completeness.
For independent and identically distributed (i.i.d.) measurements, under the regularity conditions for the Bernstein--von Mises theorem~\cite{vanderVaart1998}, the posterior becomes asymptotically Gaussian, yielding~\cite[App.~B]{rubio2018non}
\begin{equation}
    \lim_{\mu\to\infty}
    \mu\mathcal{L}_{\mathrm{MPV}}
    =
    \int \dd\theta\,
    \frac{p(\theta)}{\mathcal I(\theta)}.
    \label{eq:asymptotic-limit}
\end{equation}
For the unitary phase-estimation problem considered here, the maximum quantum FI is independent of $\theta$ [Eq.~\eqref{eq:max-qfi}], giving the asymptotic quantum CR benchmark
\begin{equation}
    \mathcal{L}_{\mathrm{CR}}
    \coloneqq
    \frac{1}{\mu\mathcal I_{\max}}.
    \label{eq:cr-benchmark}
\end{equation}
Although our adaptive protocols do not involve i.i.d. statistics, this benchmark provides a heuristic reference for assessing their asymptotic precision.

\section{Performance of adaptive Gaussian metrology}
\label{sec:results}

\begin{figure}[t!]
    \centering
    \includegraphics[width=\columnwidth, trim=0 4mm 0 0, clip]{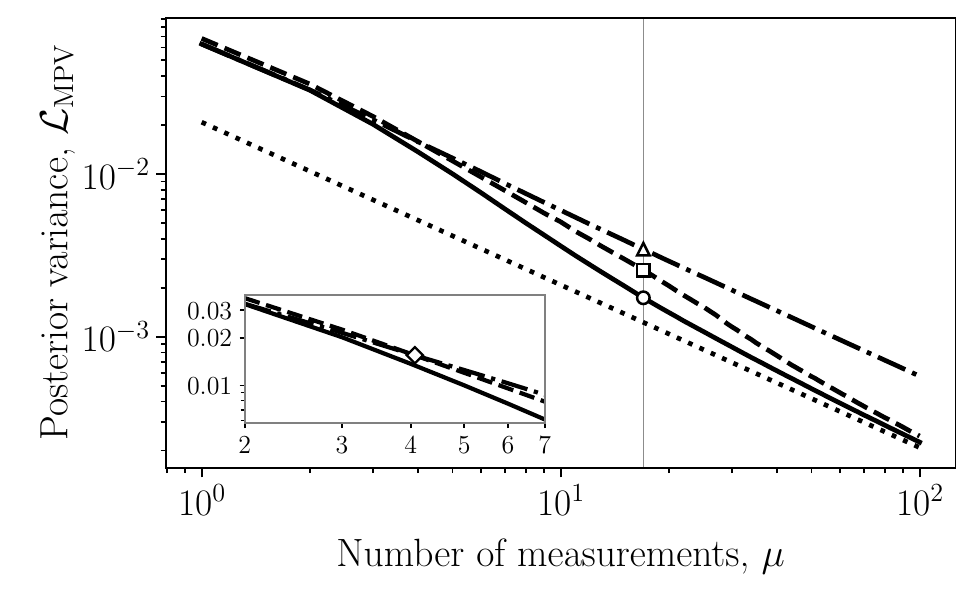}
    \caption{
    Adaptive phase estimation with single-mode Gaussian probes of fixed energy $E=2$.
    The mean posterior variance $\mathcal{L}_{\mathrm{MPV}}$ [Eq.~\eqref{eq:mpv}] is shown as a function of the number of measurements $\mu$, for adaptive homodyne detection (dashed), a fixed probe with adaptive optimal measurements (dash-dotted), and joint adaptive optimisation of probes and measurements (solid).
    The dotted line is the asymptotic quantum CR benchmark $\mathcal{L}_{\mathrm{CR}}$ [Eq.~\eqref{eq:cr-benchmark}].
    The inset highlights the first few measurements: the two strategies employing optimal quantum measurements coincide at the first shot, while the fixed-probe strategy initially outperforms homodyne detection before their ordering reverses at $\mu=4$, indicated by the diamond.
    The vertical grey line and markers identify $\mu=17$, where joint optimisation achieves its maximum relative improvement over adaptive homodyne detection (approximately $32\%$), with a corresponding improvement of $50\%$ over the fixed-probe strategy.
    The prior is uniform on $[0,\pi/2]$.
    The adaptive strategies are defined in Sec.~\ref{sec:adaptive-estimation}.
    The results are obtained from $1000$ Monte Carlo runs, using a Fock-space cutoff of $100$ for strategies involving SPM measurements.
    }
    \label{fig:multiple-rounds}
\end{figure}

We now simulate the three adaptive strategies introduced in Sec.~\ref{sec:adaptive-estimation} and assess their performance.
Fig.~\ref{fig:multiple-rounds} shows the MPV [Eq.~\eqref{eq:mpv}] as a function of the number of measurements $\mu$, together with the asymptotic quantum CR benchmark [Eq.~\eqref{eq:cr-benchmark}].
Since all strategies use the same prior and energy constraint $E=2$, their differences in performance arise from the choice of probes and measurements.

First, we observe that the relative performance can change with the number of measurements, consistent with previous results in Bayesian interferometry~\cite{Rubio_2019}.
At the first shot, both strategies employing optimal quantum measurements coincide and outperform adaptive homodyne detection.
The fixed-probe strategy initially retains an advantage over homodyne, but their ordering reverses at $\mu=4$, as highlighted in the inset.
Joint optimisation yields the lowest MPV throughout, with a strict advantage beyond the first measurement.

To quantify this advantage, we use the relative improvement defined in Eq.~\eqref{eq:relative-improvement}.
Its maximum with respect to adaptive homodyne detection occurs at $\mu=17$, where joint optimisation reduces the MPV by approximately $32\%$, compared with $50\%$ relative to the fixed-probe strategy.
Both the fully adaptive and homodyne protocols subsequently approach the asymptotic quantum CR benchmark, unlike the fixed-probe strategy.

These results quantify the precision advantage of optimal quantum measurements over experimentally accessible homodyne detection when the Gaussian probe is fully adapted in both cases.
In the regime of limited data, the improvement can be substantial even when the probe is already optimised for homodyne detection.
The comparison therefore identifies the potential precision benefit of implementing more demanding quantum measurements.

\section{Discussion and conclusions}

We have shown that jointly adapting Gaussian probes and quantum measurements can improve the precision of partially informed phase estimation with limited data.
Building upon Refs.~\cite{Rubio_2019,ricard-simon}, we have seen that, within the regime considered, joint optimisation yields the lowest uncertainty and achieves a maximum relative improvement of approximately $32\%$ over adaptive homodyne detection at $17$ measurements, with a corresponding improvement of $50\%$ over the fixed-probe strategy.

Beyond these specific results, the main contribution of this work is methodological.
The FI remains the default design metric in much of quantum metrology~\cite{DemkowiczDobrzanski2015}, partly because of the simplicity of its SLD-based recipe.
Personick's formalism, whose ingredients have been known for a while~\cite{Personick, Macieszczak_2014} but have been far less exploited, offers a way to reproduce this reasoning within a Bayesian framework.
Specifically, the Sylvester equation in Eq.~\eqref{eq:Lyapunov} plays the role of the SLD equation, while maximisation of the quantum FI over control parameters is replaced by maximisation of the prior-dependent pseudo-gain in Eq.~\eqref{eq:opt-control}.
While more general approaches exist, including the alternating optimisation of Ref.~\cite{Macieszczak_2014} and the semidefinite programming methods of Ref.~\cite{bavaresco2024designing}, the simplicity of this construction may facilitate the wider adoption of Bayesian experimental design, in line with the procedures of Refs.~\cite{glatthard2022optimal, overton2026adaptive, fess2026bayesian}
The implementation of SPM measurements remains a separate challenge.

A further feature of this framework is that averaging pure probes over the prior produces an effective mixed state, suggesting a formal connection between prior uncertainty and physical noise already noted in Ref.~\cite{Macieszczak_2014,ricard-simon}.
A related observation may apply to phase averaging in the absence of an external reference~\cite{jarzyna2012quantum,branford2021average}.
This suggests that the Bayesian paradigm could provide a common language for describing different sources of uncertainty in quantum metrology.

More technically, the optimal measurement in our calculation is generally non-Gaussian and requires a Fock-space cutoff.
Its computational cost could be reduced using the variational approach of Ref.~\cite{gandar2026closedformbayesianquantumestimation}.
Further applications could include non-Gaussian and multimode probes, as well as multiparameter estimation, where measurement incompatibility introduces additional constraints~\cite{albarelli2025measurement}.

\bigskip

\emph{Author contributions}. 
Both authors conceived the problem, interpreted the results, and wrote the manuscript.
R.R.R. performed the calculations with Gaussian states.
J.R. developed the Bayesian framework and protocol design methodology.
ChatGPT (OpenAI) and Claude (Anthropic) were used to optimise the computational efficiency of code written by the authors.
This reduced the time required to generate the plots in Fig.~\ref{fig:multiple-rounds} to a few hours on a consumer-grade laptop.
The authors verified the final results and take responsibility for them.

\bigskip

\emph{Acknowledgments}. 
We gratefully acknowledge helpful discussions with S. Morelli, D. Branford and E. Gandar. 
R. R. R. acknowledges the Government of Spain (Severo Ochoa CEX2019-000910-S and FUNQIP), Fundaci\'o Cellex, Fundaci\'o Mir-Puig, Generalitat de Catalunya (CERCA program). 
J. R. acknowledges funding from the Surrey Future Fellowship (University of Surrey), during which this work was initiated.
J. R. also acknowledges support from UK Research and Innovation, Grant No. UKRI3702, `Quantum Control and Sensing (QCS)''.

\bibliography{refs}

\appendix

\section{Gaussian probes}
\label{app:gaussian-probes}

Here we review the preparation of the Gaussian state in the main text; see Refs.~\cite{loudon2000quantum,braunstein2005quantum,weedbrook2012gaussian,serafini2017quantum,brask2022gaussianstatesoperations} for further details.

Pure single-mode Gaussian states can be prepared by applying a squeezing operator
\begin{equation}
    \hat{S}(\xi)
    =
    \exp\bigg[
        \frac{1}{2}
        (\xi^{*}\hat{a}^{2}-\xi\hat{a}^{\dagger 2})
    \bigg],
\end{equation}
and a displacement operator
\begin{equation}
    \hat{D}(\alpha)
    =
    \exp(\alpha\hat{a}^{\dagger}-\alpha^{*}\hat{a}),
\end{equation}
to the vacuum state $|0\rangle$~\cite{serafini2017quantum}.
Here, $\xi=re^{i\varphi}$ and $\alpha=|\alpha|e^{i\tau}$ denote the squeezing and displacement parameters, respectively.
The initial probe state is thus
\begin{equation}
    \ket{\psi_{\bs{y}}^{\mathrm{in}}}
    =
    \hat{D}(\alpha)\hat{S}(\xi)|0\rangle,
\end{equation}
with the squeezing strength $r$ fixed by the energy constraint in Eq.~\eqref{eq:energy-constraint}.

The phase is encoded through the rotation
\begin{equation}
    \hat{R}(\theta)
    =
    \exp(-i\theta\hat{a}^{\dagger}\hat{a}),
\end{equation}
as in Eq.~\eqref{eq:transformed-state}.
The resulting state is Gaussian in the sense that its Wigner function is Gaussian. 
It is therefore fully characterised by the first moments and covariance matrix of the quadratures
$\hat{q}=(\hat{a}+\hat{a}^{\dagger})/\sqrt{2}$ and
$\hat{p}=(\hat{a}-\hat{a}^{\dagger})/(i\sqrt{2})$.
These are given in Eqs.~(\ref{eq:first moment single mode states_2}--\ref{eq:covariance-componentes}), with covariance entries
\begin{equation}
    V_{ij}
    =
    \frac{1}{2}
    \langle
        \{\hat{\bs{r}}_i-\langle\hat{\bs{r}}_i\rangle,
          \hat{\bs{r}}_j-\langle\hat{\bs{r}}_j\rangle\}
    \rangle,
\end{equation}
where $\hat{\bs{r}}=(\hat{q},\hat{p})^T$ and
$\{A,B\}=AB+BA$.

\section{Loss function for phase estimation}
\label{app:loss-function}

Since phase is a circular parameter, the loss function $\ell(\tilde{\theta},\theta)$ must respect the periodicity $\theta' = \theta+2\pi$.
A natural choice is the distance between two points on the unit circle, which assigns zero loss to equivalent phases.
Representing each phase by the unit vector $\bs{u}(\theta) = (\cos\theta, \sin\theta)^{\transpose}$, we obtain
\begin{align}
    \ell(\tilde{\theta},\theta)
    &= \|\bs{u}(\tilde{\theta})-\bs{u}(\theta)\|^2 \nonumber\\
    &= 2[1-\cos(\tilde{\theta}-\theta)] \nonumber\\
    &= 4\sin^2\left(\frac{\tilde{\theta}-\theta}{2}\right),
\end{align}
where $\|\cdot\|$ is the Euclidean norm.
This gives the sinusoidal loss discussed in the main text.

\end{document}